\documentclass[]{mn2e}
\usepackage{epsfig}
\newif\ifAMStwofonts
\newcommand{\ltappeq}{\raisebox{-0.6ex}{$\,\stackrel
        {\raisebox{-.2ex}{$\textstyle <$}}{\sim}\,$}}

\title{Is the slope of the intrinsic Baldwin effect constant?}
\author[M.R. Goad et al.]
       {M.R. Goad$^{1}$, K.T. Korista$^{2}$, C. Knigge$^{1}$\\
        (1) University of Southampton, (2) Western Michigan University}

\date{Received 2004 Feb 17}

\pagerange{\pageref{firstpage}--\pageref{lastpage}}
\pubyear{2002}
\begin{document}

\maketitle

\label{firstpage}

\begin{abstract}

We investigate the relationship between emission-line strength and
continuum luminosity in the best-studied nearby Seyfert~1 galaxy
NGC~5548. Our analysis of 13~years of ground-based optical monitoring
data reveals significant year-to-year variations in the observed
H$\beta$ emission-line response in this source. More specifically, we
confirm the result of Gilbert and Peterson (2003) of a non-linear
relationship between the continuum and H$\beta$ emission-line fluxes.
Furthermore, we show that the slope of this relation is not constant,
but rather decreases as the continuum flux increases. Both effects are
consistent with photoionisation model predictions of a
luminosity-dependent response in this line.

\end{abstract}


\section{Introduction}
% AGNs. - short intro.

Observations of correlated continuum and broad emission-line
variations in Active Galactic Nuclei (AGN) provide powerful
diagnostics of the structure and physical conditions within the
spatially unresolved broad emission-line region (BLR) and of the
origin and shape of the unobservable ionising continuum.

A well-established correlation is the observed decrease
in broad emission-line equivalent width with increasing continuum
level for AGN spanning a broad range in continuum luminosity; this is
the so-called global ``Baldwin effect'' (Baldwin 1977; Osmer, Porter and Green
1994).  Originally observed in the strongest UV resonance
lines (e.g. CIV$\lambda 1549$ and Ly$\alpha$), a global Baldwin
effect has now been found for most of the strong UV emission-lines.
%apart from NV$\lambda1240$.\footnote{Dietrich et al. suggest that
%abundance enhancements at higher z, may obscure the relationship for
%this line.} 
Interestingly, evidence for a similar effect in the broad optical
hydrogen recombination lines remains weak.
%\footnote{The absence of a
%global Baldwin effect in the optical recombination lines may indicate
%that the strength of these lines scales simply with source luminosity
%and in such a way that their line strengths simply reflect the number
%of hydrogen ionising photons. If correct, then this necessarily
%implies that these lines arise in material which is ionisation
%bounded.}  (Dietrich et al. 2003).  
%While the Baldwin effect is well-substantiated, its origin was, until
%recently, poorly understood. Proposed explanations include : (i) a
%softening of the ionising continuum, or a decrease in the gas covering
%fraction, with increasing source luminosity or (ii) evolution of the
%BLR with cosmic time. However, 

In a recent study of composite quasar spectra spanning a broad range
in continuum luminosity and redshift, Dietrich et~al.\ (2003) found no
evidence for evolution of the Baldwin effect with cosmic time. Instead
they suggested that the effect is a consequence of
luminosity-dependent spectral variations, in the sense that the
ionising continuum becomes softer for higher continuum luminosities.
This agrees well with the theoretical work by Korista et~al.\ (1998),
who showed that if the gas metallicity and continuum spectral energy
distribution are related statistically to the quasar luminosity, then
so will be the emission-line equivalent widths, in a manner described
by the Baldwin effect.

Formally, the relationship between the continuum luminosity
($L_{cont}$) and broad emission-line luminosity ($L_{line}$) may be
represented by a single powerlaw, such that

\begin{equation}
 L_{line} \propto L_{cont}^{\alpha} \, .
\end{equation}

\noindent In terms of the line equivalent width, the Baldwin relation
is then given by

\begin{equation}
 EW_{line} \propto L_{cont}^{\beta} \, ,
\end{equation}

\noindent where $\beta=\alpha-1$. The measured values of $\alpha$ are
typically $<1$, for example, $\alpha\approx0.83$ for CIV and
$\alpha\approx0.88$ for Ly$\alpha$ (see e.g. Kinney, Rivolo \&
Koratkar 1990, Pogge \& Peterson 1992), with corresponding slopes for
the Baldwin relation of $\beta=-0.17$, and $\beta=-0.12$ respectively.

\begin{figure*}
%\vspace*{474pt}
\epsfig{width=6.8in,angle=0,file=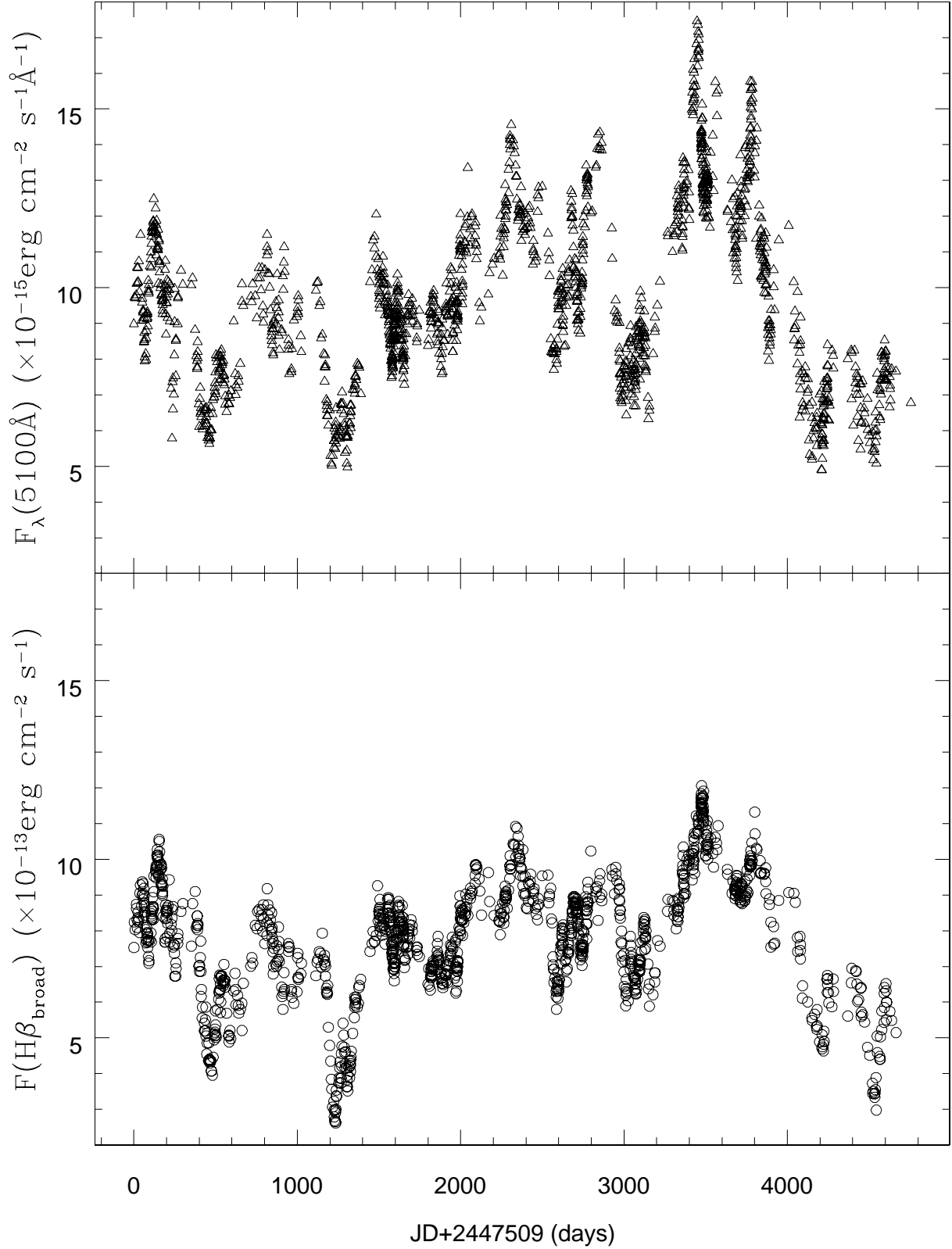}
\caption{(Upper panel)$F(\lambda5100\AA)$ continuum light curve 
(observed frame) for the full 13~yr AGNWatch monitoring campaign
of NGC~5548. (Lower panel) The corresponding H$\beta$ emission-line 
light curve.}
\end{figure*}

\begin{figure*}
%\vspace*{474pt}
\epsfig{width=3.4in,angle=0,file=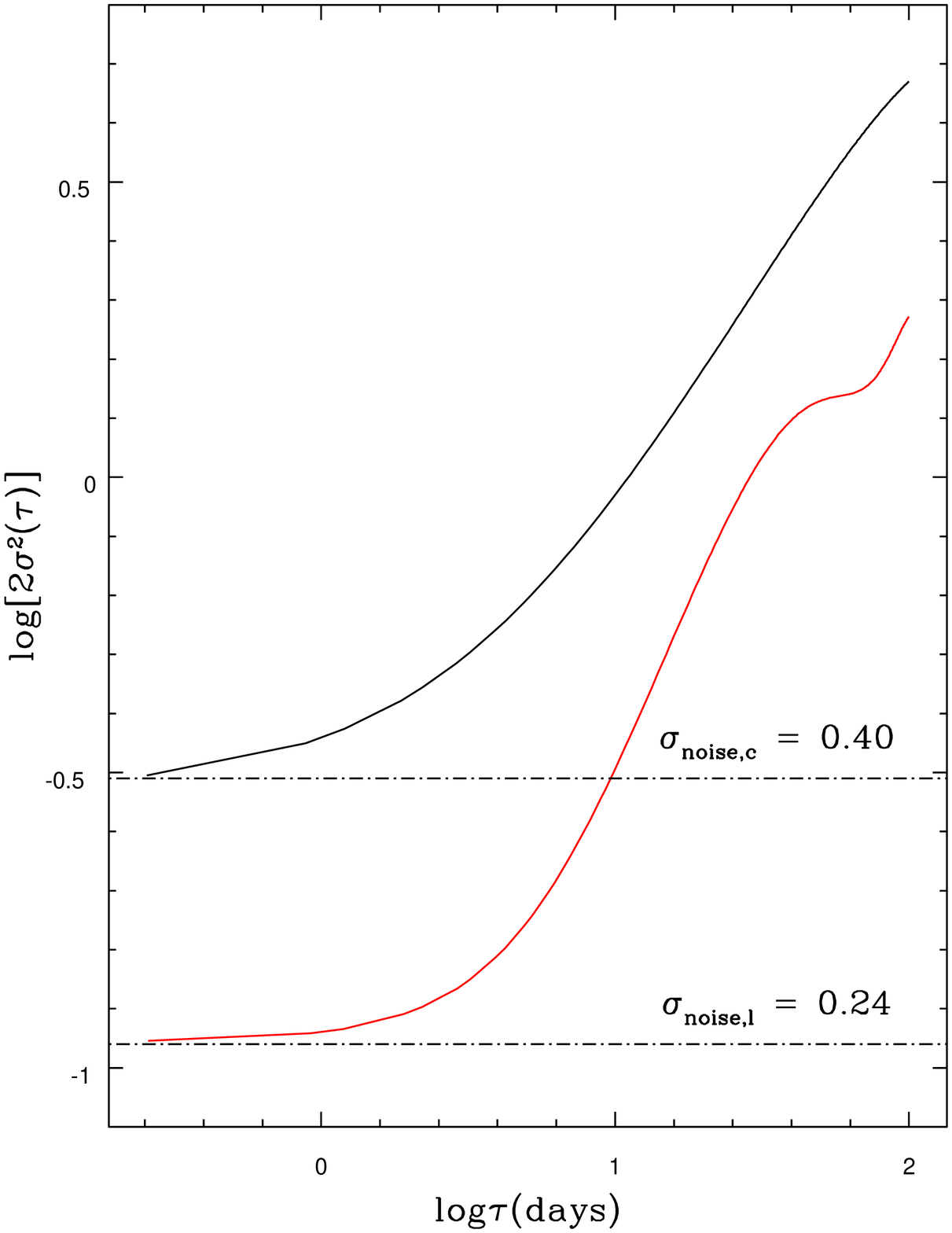}
\caption{Computed structure functions for the continuum (upper curve)
and H$\beta$ light curves. Dashed lines indicate the noise level in
each of the light curves.}
\end{figure*}

Superposed on the global Baldwin relation, is a second relation whose
slope reflects the emission-line response to variations in the
ionising continuum within an individual source.  Now commonly referred
to as the ``intrinsic Baldwin effect'', this phenomenon accounts for
at least some of the scatter seen in the global Baldwin relation.
However, the intrinsic Baldwin relation also shows considerable
scatter, whose origin is largely attributable to
continuum--emission-line time-delay effects (reverberation) within the
physically extended broad emission-line region (see e.g. Krolik
et~al.~1991, Pogge and Peterson 1992, Peterson et~al.~2002).

Formally, the relationship between the continuum flux $F_{\rm cont}$
and emission-line flux $F_{line}$ within a single source can also
be represented by a single powerlaw, such that

\begin{equation}
 F_{line} \propto F_{cont}^{\alpha} \, ,
\end{equation}

\noindent where in this case $\alpha$ is a measure of the
instantaneous emission-line response to changes in the ionising
continuum flux, the {\em responsivity\/} of the gas. In terms of the
line equivalent width, the intrinsic Baldwin relation is then given by

\begin{equation}
 EW_{line} \propto F_{cont}^{\beta} \, .
\end{equation}

\noindent where again $\beta=\alpha-1$.

While a global Baldwin relation has yet to be seen in the optical
recombination lines, there is no physical reason why an individual
source should not display an intrinsic Baldwin effect in these lines.
That is to say, the broad range in physical conditions extant within
the BLR are such that case~B recombination, for which $\alpha\equiv
1$, does not apply.  Indeed, Gilbert \& Peterson (2003) recently
reported just such an effect for the broad H$\beta$ emission-line in a
comprehensive study of 10 years of ground-based optical monitoring
data on the nearby Seyfert~1 galaxy NGC~5548, taken as part of the
AGNWatch collaboration. Interestingly, Gilbert \& Peterson noted
(their Figure~4) that during a low continuum flux state (year 4 of the
monitoring campaign), the relationship between the continuum and
H$\beta$ emission-line flux, appeared to change.  However, since there
was only {\em one\/} low flux state, they were unable to determine
whether this change was a result of flux-dependent effects
(e.g. photoionisation) or time-dependent effects (e.g. a
renormalization of the intrinsic relation due to a change in
composition, and/or distribution of the line-emitting gas).

While variations in the emission-line response with continuum state
have long been predicted by photoionisation calculations (e.g.
O'Brien et~al. 1995; Goad 1995; Korista \& Goad 2004), they have so
far never been reported in the observations. Here, we take a detailed
look at all 13~years of optical continuum and emission-line data
available for NGC~5548, including two new low-flux states. Using this
unique dataset we have found the first observational evidence for
luminosity-dependent variations in the emission-line response.  These
luminosity-dependent variations are not only apparent in the whole
dataset (reinforced by the addition of new low continuum flux data
from years 12 and 13), but are even visible within individual
observing seasons, being particularly prominent during those seasons
displaying both high and low continuum flux states (e.g. year 4, see
\S3).  Since the observed variations can occur on timescales of less
than one year, they most likely represent a response to luminosity
variations rather than to structural changes in the emission-line
region.

%%%%%%%%%%%%%%%%%%%%%%%%%%%%%%%%%%%%%%%%%%%%%%%%%%%%%%%%%%%%%%%%%%%%%%%%%%%%

\section{13~yrs of optical data}

The bright ($\lambda\/L(\lambda5100\rm{\AA\/})\sim
10^{43.5}$erg~s$^{-1}$), nearby (z=0.017), Seyfert~1 galaxy NGC~5548
is perhaps the best-studied of all AGN at optical wavelengths.  This
galaxy has been the subject of a concerted monitoring campaign by the
AGNWatch collaboration since 1990, having been observed over 1500
times with a mean sampling interval of $\sim3$~days (see
e.g. Peterson et~al.~[2002] for a comprehensive review).

Here we analyse the optical continuum and H$\beta$ emission-line data
for NGC~5548 in order to determine whether the H$\beta$ emission-line
response for this source is sensitive to the continuum level
throughout the 13~year duration of this campaign. For homogeneity, we
use only data taken with the Ohio-State CCD spectrograph on the 1.8~m
Perkins Telescope at the Lowell Observatory in Flagstaff Arizona.  For
a detailed review of the reduction procedure we refer the reader to
Peterson et~al.\ (2002) and references therein. In brief, the
continuum flux was measured in a 10~\AA\/ wide bin centred at
$\lambda$5100~\AA\/ in the rest frame of the galaxy. The H$\beta$
emission-line flux was then determined by integrating the flux over a
linear fit to the background continuum between $\lambda$4710~\AA\/ and
$\lambda$5100~\AA\/ again in the rest-frame of the galaxy.  Figure~1
shows the full 13 years of continuum and line flux measurements
(observed frame) taken as part of the AGNWatch campaign.

\subsection{Background removal}

Both continuum and emission-line fluxes are contaminated by nuisance
background components which must be removed.  The continuum flux
measurements are contaminated by a constant background contribution by
starlight from the host galaxy. In addition, the broad H$\beta$
emission-line flux is contaminated by a non-variable narrow-line
component.  For each component we adopt the canonical values used by
Gilbert \& Peterson (2003) in their fit to the first 10 years of the
AGNWatch data ie.  $F_{\rm
galaxy}=3.37\pm0.54\times10^{-15}$~erg~cm$^{-2}$~s$^{-1}$~\AA$^{-1}$
for the continuum flux (Romanishin et~al. 1995) and $F(H\beta_{\rm
narrow})=6.7\pm0.6\times$10$^{-14}$~erg~cm$^{-2}$~s~$^{-1}$.  

\begin{figure*}
%\vspace*{474pt}
\epsfig{width=3.4in,angle=0,file=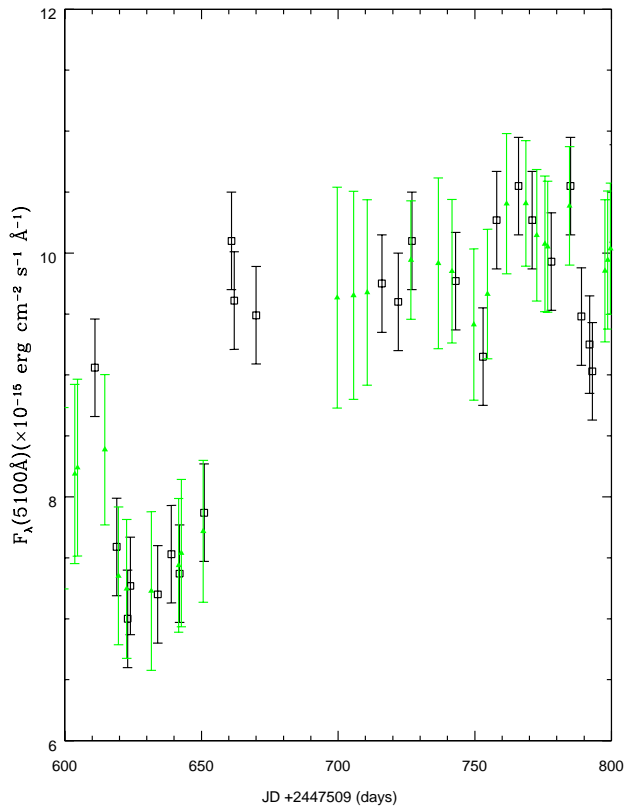}
\caption{A section of the original continuum light curve (open squares) 
and the reconstructed continuum flux points (filled triangles) together with
their associated uncertainties.}
\end{figure*}

\begin{figure*}
%\vspace*{474pt}
\epsfig{width=6.8in,angle=0,file=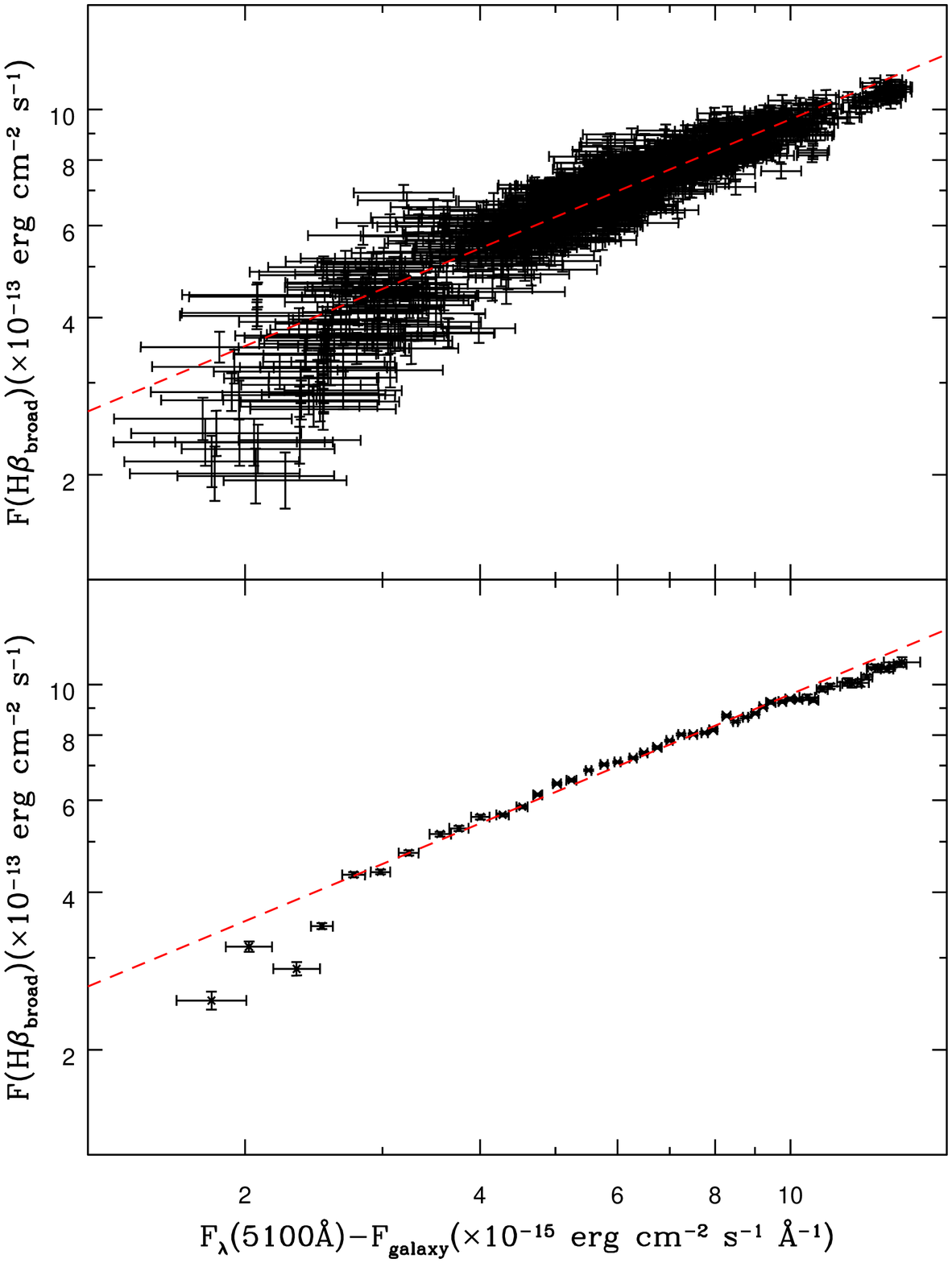}
\caption{(Upper panel) Best-fit slope (dashed line) to 
the full 13~yrs of optical data.  (Lower panel) As above, only this
time we show the data binned in 0.25 intervals in continuum flux (see
text for details).}
\end{figure*}

\subsection{Delay removal and error estimates from structure function
analysis}

The continuum and H$\beta$ light curves are highly correlated, but the
emission line variations are delayed with respect to the continuum
(Fig~1). Physically, these delays arise because the continuum is
produced close to the central engine of an AGN whereas the
line-forming region is located at larger distances, in the spatially
extended BLR. Given that the line emission is ultimately powered by
the continuum incident on the BLR, the lines thus respond to changes
in the continuum with a time-delay, $\tau$, that is a measure of the
luminosity-weighted radius, $R_{\rm BLR}$, of the BLR, ie. $\tau \simeq
R_{\rm BLR} /c$.

When determining emission-line responsivities, these delays are
contaminants and must be corrected for. This is because the line
responsivity is defined as the instantaneous emission-line response to
short-timescale small amplitude changes in the continuum flux incident
on the BLR at the same time ie. $\alpha=d\log F_{line}/d\log
F_{cont}$.  In practice, we therefore shift the emission-line data for
each year by their respective delays (as calculated from the centroids
of the cross-correlation function for each of the 13 years of data,
see e.g. Peterson~et~al.\ 2002). We then estimate the continuum flux at
each (shifted) emission line epoch as the weighted average of the two
bracketing continuum points. 

The appropriate weight for each point is calculated from the first
order structure function derived from the continuum light curve.
Detailed descriptions of structure function analysis may be found in
Kawaguchi et~al.\ (1998) or Paltani (1999) and references
therein. Briefly, the first order structure function for a series of
flux measurements $f(t_{i})$, $i=1,N$ is defined as

\begin{equation}
S(\tau)=\frac{1}{N(\tau)}\sum_{i<j}[f(t_{i})-f(t_{j})]^{2} \, ,
\end{equation}

\noindent where the sum is over all pairs of points for which
$t_{j}-t_{i}=\tau$, and $N(\tau)$ is the total number of pairs of
points. It is straightforward to show that $S(\tau)$ is a measure of
(twice) the variance of the light curve on timescale $\tau$,
i.e. $S(\tau) \simeq 2\sigma^{2}(\tau)$. Correspondingly, structure
functions are usually characterised by two flat sections with
amplitudes of twice the total variance of the data, $2\sigma^{2}$, on
the longest timescales and twice the noise variance due to observational 
errors, $2\sigma_{noise}^{2}$, on the shortest timescales. These flat
sections are typically joined by by a rising power law on intermediate
timescales. 

Figure~2 shows the structure functions for both the continuum (upper
curve) and emission-line (lower curve) light curves for timescales of
$<$500~days. Given a pair of continuum points, the continuum structure
function tells us immediately the correct weight to assign each point
in estimating the continuum flux at any time, $t$, between them. More
specifically, each of the two points bracketing $t$ is assigned the
usual inverse variance weight $2/S(\tau) \simeq 1/\sigma^{2}(\tau)$,
where $\tau = t_i - t$ and $t_i$ is the time associated with each
point.

We also use the structure functions to estimate the errors on both
emission-line and continuum flux estimates. Figure~2 shows that the
structure functions are flat on the shortest timescales ($\tau
\ltappeq$~1~day). The corresponding estimates of the 
instrumental uncertainties are $\sigma_{noise,c}=0.4\times
10^{-15}$~erg~cm$^{-2}$~s$^{-1}$~\AA$^{-1}$\ for the continuum and
$\sigma_{noise,l}=0.24\times 10^{-13}$~erg~cm$^{-2}$~s$^{-1}$\ for the
line. These respectively correspond to fractional errors of $\sim3$\%
and $\sim2$\% at high continuum and line flux levels, rising to
$\sim25$\% and $\sim15$\% at low continuum and line flux levels. Since
shifting the line data points leaves their flux values unaltered, we
simply take the error on the line fluxes to be $\sigma_l =
\sigma_{noise,l}$. By contrast, the reconstructed continuum points are
two-point weighted averages of the bracketing data values. We
therefore assign errors corresponding to the standard deviation
estimated from these data values,
\begin{equation}
\sigma_c^2 = \frac{2}{\sum_{i=1}^{2}\frac{1}{\sigma_(\tau_{i})^{2}}}.
\end{equation}
The factor two in the numerator is needed because we want the standard
deviation of the two bracketing points, not the error on the
mean. This ensures that the minimum error associated with a
reconstructed point is $\sigma_{noise,c}$ (rather than
$\sigma_{noise,c}/\sqrt{2}$). 

Figure~3 shows the input continuum (open squares), reconstructed
continuum (filled triangles) and their associated errors for a small
section of the 13~year light curve. Note that in filling data gaps of
less than a day, our reconstruction assigns uncertainties comparable
to those on the original data points (i.e., $\sigma_{noise,c}$). Where
gaps are longer (e.g., boundaries between observing seasons) the
uncertainties on the reconstructed points in the gaps increase with
increasing distance from the nearest bracketing point. Both types of
limiting behaviour are sensible.

%We finally note that our approach to continuum reconstruction and
%error estimation differs from that of Gilbert \& Peterson
%(2003). First, they adopted a fractional
%error on the integrated continuum and line fluxes of 2.5\% as a
%reasonable estimate of the likely uncertainty on these measurements
%Gilbert \& Peterson reconstructed continuum fluxes
%corresponding to (shifted) emission line epochs by direct 
%linear interpolation on bracketing continuum data points. Second, 
%they determined the uncertainty associated with the reconstructed 
%points by adding the errors in the neighbouring flux
%points in quadrature. Our own analysis suggests that the
%uncertainties\footnote{By analysing the flux differences between
%neighbouring pairs of points Gilbert and Peterson concluded that flux
%differences on short timescales (a few days) are purely stochastic
%rather than real differences. They therefore adopted a fractional
%error on the integrated continuum and line fluxes of 2.5\% as a
%reasonable estimate of the likely uncertainty on these measurements.}
%they assumed for individual points and for the reconstructed continuum
%points are likely too small.  Here we adopt a more conservative
%approach using a 
%structure function analysis (see e.g. Kawaguchi 
%et~al.~1998, Paltani 1999) of the continuum and emission-line data to
%estimate the likely measurement uncertainties in both light curves,
%and to estimate the location and uncertainty on reconstructed
%continuum flux points.

\section{Results}

Figure~4 (upper panel) shows the $\lambda$5100\AA\ continuum flux
versus H$\beta$ emission-line flux after correcting for contaminating
background components and the yearly continuum--emission-line time-delays.
Also shown (dashed line) is a linear least squares fit to the data
accounting for errors in both coordinates, and assuming that the data
follow the relation

\begin{equation}
\log F({\rm H}\beta_{\rm broad}) = \log A + \alpha \log[F_{\lambda}(5100\AA)-
F_{galaxy}] \, .
\end{equation}

\noindent The best-fit slope $\alpha$, and 
1$\sigma$ uncertainty (estimated using bootstrap re-sampling) together
with the mean continuum flux (unweighted) and root mean square error
are given in Table~1.  Our estimated slope and 1$\sigma$ uncertainty
for the 13~year campaign, $0.621\pm0.019$ (Table 1), is marginally
smaller (but still within the estimated uncertainties) than that found
by Gilbert and Peterson (2003) for the first 10 years of data. We
therefore confirm Gilbert and Peterson's finding of an intrinsic
Baldwin effect for this line with slope $\beta\approx-0.4$.

Figure~4 -- lower panel -- shows the same relation plotted
against the binned continuum data. The continuum data were binned in
0.25 intervals in continuum flux with individual points in each bin
weighted according to the uncertainties on the reconstructed continuum
flux. The binned continuum data indicate evidence not only for a
steepening of the observed relation at low continuum flux levels, as
previously noted by Gilbert and Peterson (2003), but also for a
flattening of the relation at the highest continuum fluxes.

In order to discriminate between temporal changes in the structure of
the BLR and a luminosity-dependent emission-line response, we have
also applied the same fitting routine to each year of data separately.
Figure~5 shows the best fit slopes (solid lines) for each of the
13~years of data. To guide the eye, we also show the best fit slope
(dashed line) determined for the full 13~year campaign.  The measured
slopes and their 1$\sigma$ uncertainties (again determined using
bootstrap re-sampling) are given in Table~1.

\begin{figure*}
%\vspace*{474pt}
\epsfig{width=5.5in,angle=270,file=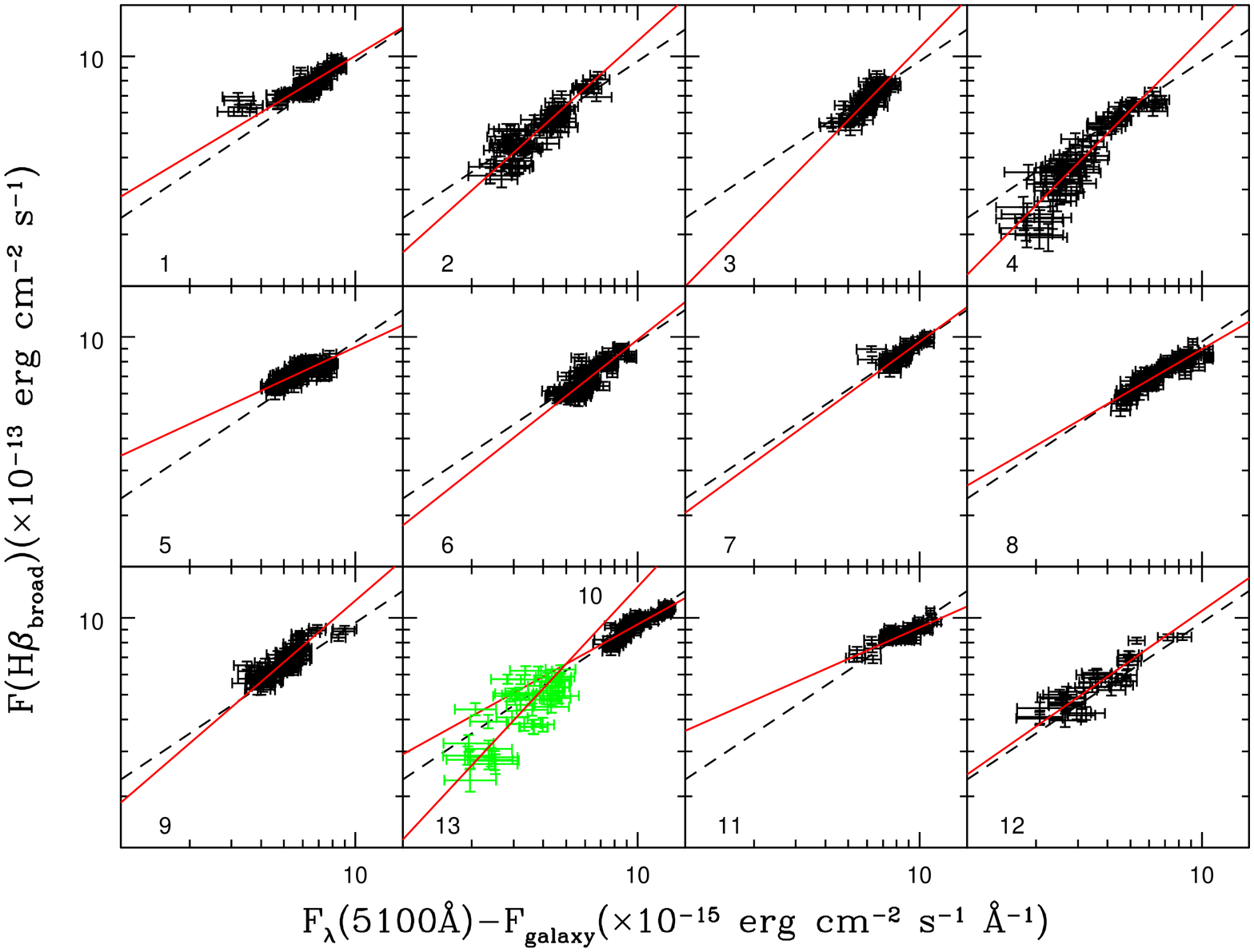}
\caption{Best-fit line responsivities, $\alpha$, 
for each of the 13~years of optical data. 
To guide the eye, we also show the
best-fit line responsivity (dashed line) for the complete 
data set. For ease of display, 
panel 10 contains fits to year 10 and year 13.}
\end{figure*}

\begin{figure*}
%\vspace*{474pt}
\epsfig{width=3.4in,angle=0,file=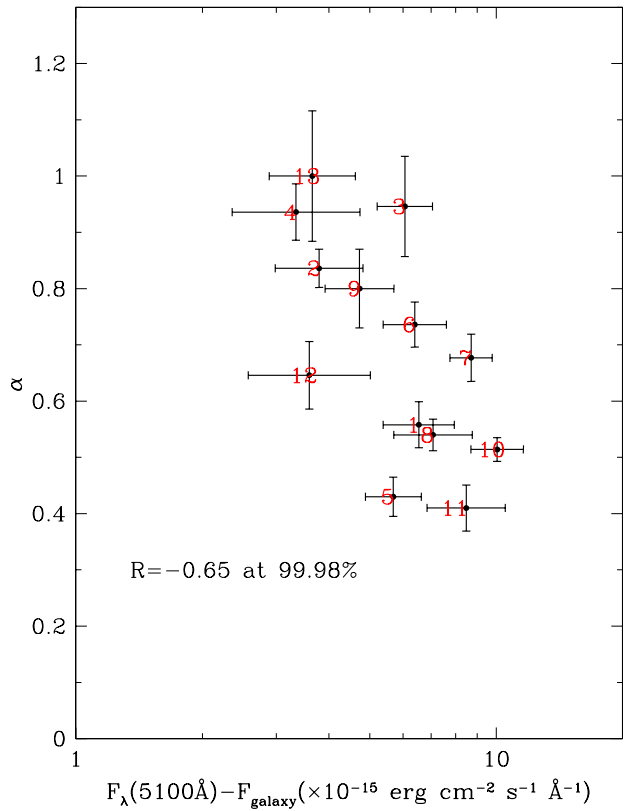}
\caption{The measured correlation between the line responsivity $\alpha$, and
continuum flux. The numbers indicate the corresponding observing
season.  A Pearson's' rank correlation shows that $\alpha$ is
anti-correlated with continuum flux at better than $3\sigma$.}
\end{figure*}

\begin{table*}
%\vspace{124pt}
\centering 
\caption{The measured H$\beta$ emission-line responsivities,  $\alpha$, 
and their 1$\sigma$ uncertainties for combined and yearly data
sets. Also shown are the corresponding values for the intrinsic
Baldwin relation $\beta$. Quoted flux values are the unweighted means
and rms error in units of
$10^{-15}$~erg~cm$^{-2}$~s$^{-1}$~\AA$^{-1}$.}
\begin{tabular}{@{}rrrrrr@{}} Year & $\alpha$ & $\beta=\alpha-1$ & $\sigma_{\alpha}$ & 
$<F_{cont}>$ & $\sigma_{<F_{cont}>}$ \\ & & &$\sigma_{\beta}$ & ($\times10^{-15}$) & \\
\hline
all years & 0.621 &  $-$0.379  & 0.019 &  6.33 & 2.41 \\
1      	  & 0.558 &  $-$0.442  & 0.041 &  6.54  & 1.27\\
2 	  & 0.836 &  $-$0.164  & 0.034 &  3.79  & 0.91\\
3 	  & 0.946 &  $-$0.054  & 0.089 &  6.06  & 0.92\\
4  	  & 0.936 &  $-$0.064  & 0.050 &  3.34  & 1.17\\
5  	  & 0.430 &  $-$0.570  & 0.035 &  5.69  & 0.87\\
6 	  & 0.736 &  $-$0.264  & 0.040 &  6.40  & 1.11\\
7  	  & 0.677 &  $-$0.323  & 0.042 &  8.71  & 1.01\\
8  	  & 0.540 &  $-$0.460  & 0.028 &  7.07  & 1.52\\
9  	  & 0.800 &  $-$0.200  & 0.070 &  4.73  & 0.89\\
10 	  & 0.514 &  $-$0.486  & 0.021 & 10.05  & 1.44\\
11 	  & 0.410 &  $-$0.590  & 0.041 &  8.48  & 1.82\\
12 	  & 0.646 &  $-$0.354  & 0.060 &  3.59  & 1.20\\
13 	  & 1.000 &  0.000     & 0.116 &  3.65  & 0.86\\
\hline
\end{tabular}
\end{table*}

\section{Discussion}

Figure~5 and Table~1 demonstrate that the H$\beta$ emission-line
responsivity, $\alpha$, shows significant variations from one
observing season to the next. To quantify this further, we show in
Figure~6 the mean continuum flux for each observing season versus
$\alpha$ for each of the individual years. A Pearson's rank
correlation shows that $\alpha$ is anti-correlated with the
optical continuum luminosity with slope $-0.65$ at better than
$3\sigma$, strongly suggesting that the continuum luminosity and
H$\beta$ line responsivity are intimately connected.  Further
corroborating evidence for such a relation can be seen in panel~4 of
Figure~5. Season~4 displays the largest range in continuum variation
of any of the 13 seasons. It also shows a marked increase in slope
with decreasing continuum level. A similar trend though not as
notable, is also seen in year~2. Such short timescale variations in
the line-responsivity are unlikely to be due to dynamical effects
within the BLR. The dynamical timescale $\tau_{dyn}$ for the BLR is
given by $\tau_{dyn}\approx R/v_{\rm fwhm}$, where $R$ is the
luminosity-weighted ``size'' of the BLR, and $v_{\rm fwhm}$ is the
full-width at half-maximum of the root mean square emission-line
profile. Adopting the mean BLR size of $\sim20$~lt-days and mean FWHM
of the H$\beta$ profile of $\sim$5000~km~s$^{-1}$), $\tau_{\rm
dyn}\approx3$~yrs  for this source, far longer
than any single observing season.

Given the relatively short timescales over which changes in the
emission-line responsivity occur, our preferred interpretation is that
variations in $\alpha$ are likely related to changes in continuum flux
only. While there are clearly points which do not obey this general
trend, e.g.  year~7, we suspect that this merely reflects our
inability to adequately account for reverberation effects within the
spatially extended BLR\footnote{The lag is a one number estimate of
the luminosity-weighted ``size'' of a region which is spatially
extended.}.  For example, we expect the largest discrepancies from the
overall trend in Figure~6 whenever a low continuum state directly
follows a high one. This is due to a residual contribution to the
overall response in the low state from emission-line gas in the outer
BLR. The time delay associated with this region is longer than the
luminosity-weighted average, so it may still be responding to the
prior high-state continuum flux. Residual reverberation effects such
as these most likely account for the comparatively weak response of
the line during year 12, a low continuum state which follows two
previous high continuum states (years 10 and 11).

%%%%%%%%%%%%%%%%%%%%%%%%%%%%%%%%%%%%%%%%%%%%%%%%%%%%%%%%%%%%%%%%%%%%%%%%

We note that the measured line-responsivities presented here indicate
the relationship between the integrated H$\beta$ emission-line flux
and $\lambda5100$\AA\ continuum flux. However, to ascertain the
mechanism behind this relation it is necessary to relate the
emission-line flux to the ionizing continuum flux.  Since we cannot
observe the ionising continuum directly, we adopt the
$\lambda1350$\AA\ continuum flux as a reasonable surrogate.
Contemporaneous UV/optical continuum data (Gilbert \& Peterson 2003) reveal
a relationship of the form

\begin{equation}  
F(\lambda5100\AA) \propto F(\lambda1350)^{2/3} \, .
\end{equation}

\noindent Thus the slopes relative to the driving ionizing continuum 
may be up to a factor of 2/3 smaller than those presented here, 
ranging from $\sim0.27$ at high continuum
flux levels, to $\sim0.67$ at low continuum flux levels.

An enhanced line-response at low incident continuum levels and a
reduced line-response at high continuum levels matches well the
predicted temporal behaviour of the optical recombination lines in
recent detailed photoionisation calculations of the BLR in NGC~5548
(Korista and Goad 2004). The origin of this behaviour is thought to
arise from the strong dependence of these lines' emissivities on the
incident ionising photon flux which results in a marked reduction in
their responsivity with decreasing distance from the central ionising
continuum source. Their responsivities should therefore display
temporal variations due to large changes in the continuum luminosity,
with their responsivities expected to be anti-correlated with the
continuum level. This is in stark contrast to the behaviour of the
high ionization lines (HILs) whose lines responsivities are more
strongly correlated with the overall ionization state of the gas than
on the incident ionising continuum flux.  Since the line responsivity
$\alpha$ is on average smaller for the optical recombination lines
than for the HILs, we expect a stronger intrinsic Baldwin effect for
these lines than for the HILs, as is generally observed (see
e.g. Krolik et~al.\ 1991; Pogge \& Peterson 1992; Dietrich et~al.\
2003).

\section{Conclusions}

Our analysis of $\sim$13 years of optical continuum and H$\beta$
emission-line data for NGC~5548 shows that the H$\beta$ emission-line
responsivity and hence slope of the intrinsic Baldwin effect $\beta$
displays significant variation on timescales $\sim1$~yr. The line
responsivity is generally anti-correlated with continuum level such
that in high continuum states the emission-line response is weaker on
average. Conversely in low continuum states the responsivity is
stronger than on average. This behaviour is consistent with predictions
of photoionisation models.

\section{Acknowledgements}
We would like to thank the referee, Ari Laor, for helpful comments
leading to clarification of the key points presented in this work.
MRG would also like to thank the generous hospitality of the family
Korista during the initial stages of this work.

\label{lastpage} 

\end{document}